\documentclass[12pt]{article}
\usepackage{pifont}
\usepackage{amsmath}
\allowdisplaybreaks[4]
\usepackage{amssymb,float}
\usepackage{amsthm,graphicx,pifont}
\usepackage{graphicx,psfrag,float,subfigure}
\usepackage{caption,enumerate}
\usepackage[scale=0.8,a4paper]{geometry}
\usepackage[colorlinks]{hyperref}
\usepackage{amsthm,amsmath,amssymb}
\usepackage{mathrsfs}
\usepackage{amsmath,amssymb}
\usepackage{algorithm,algorithmicx,algpseudocode}

\title{Semitotal domination in unit disk graphs}
\author{Mingjun Liu$^{a}$, {Weiping Shang$^{b}$\thanks{e-mail: mingjunliu123@163.com, shangwp@zzu.edu.cn.}}\\
{\small $^{a}$School of Statistics and Big Data, Zhengzhou College of Finance and Economics, China}\\
{\small $^{b}$School of Mathematics and Statistics, Zhengzhou University, China}\\
}
\date{}\makeatother

\begin{document}
\maketitle

\begin{abstract}
A set $S \subseteq V$ is called a {\em semitotal dominating set} of $G=(V,E)$ if every vertex in $V \setminus S$ is adjacent
to at least one vertex in $S$, and every vertex in $S$ is within distance 2 of another vertex in $S$.
The corresponding decision problem is NP-complete even for unit disk graphs. In this paper,
we present a 5-factor approximation algorithm for the Minimum Semitotal Domination problem on unit disk graphs in the graph-based input model.
The algorithm processes the layers of a Breadth-First-Search tree and constructs a maximal independent set whose vertices satisfy the semitotal condition.
For a graph with $n$ vertices and $m$ edges, the algorithm runs in $O(n + m)$ time, and hence in $O(n^2)$ time in the worst case. This improves
the previously known 5.75-approximation algorithm with $O(n^3)$ running time.
\end{abstract}


\medskip


\section{Introduction}

In this paper, we consider only simple undirected connected graphs with at least two vertices.
For standard graph-theoretical notation and terminology, we refer the reader to
\cite{BM08}. Let $G=(V,E)$ be a graph with $n=|V|$ vertices and  $m=|E|$ edges.
For a vertex $v\in V$, let $N(v)=\{ u\in V: uv\in E\}$ and $N[v]=N(v)\cup \{ v\}$ denote its
open and closed neighborhoods, respectively. The \textit{degree} of vertex $v \in V$ is $d(v) = |N(v)|$.
For $S\subseteq V$, let $N(S)$ denote the neighbor set of vertices in $S$.
For vertices $u, v \in V$, the distance between them in $G$ is denoted by $d(u, v)$.

A set $ I\subseteq V$ is called an {\em independent set} (IS) of
$G$ if all vertices in $I$ are pairwise non-adjacent, and it is
further called a {\em maximal independent set} (MIS) if each vertex
$V\setminus I$ is adjacent to at least one vertex in $I$.

A set $D \subseteq V$ is called a {\em dominating set} (DS) of $G$ if every vertex in $V \setminus D$ is adjacent
to at least one vertex in $D$. The minimum cardinality of a dominating set is the {\em domination number} $\gamma(G)$.
A DS is called a {\em total dominating set} (TDS) if it also induces a subgraph with no isolated vertex.
The minimum cardinality of a total dominating set is the {\em total domination number} $\gamma_t(G)$.

A set $S \subseteq V$ is called a {\em semitotal dominating set} (STDS) of $G$  if it
is a dominating set and every vertex in $S$ is within distance 2 of another vertex in $S$.
The minimum cardinality of such a set is the {\em semitotal domination number} $\gamma_{t2}(G)$.

Clearly, every semitotal dominating set is also a dominating set, and every total dominating set is a semitotal dominating set.
Consequently, the semitotal domination number of a connected graph $G$ with $n\geq 2$ satisfies
$\gamma(G) \leq \gamma_{t2}(G) \leq \gamma_t(G)$.

A graph $G$ is a \textit{unit disk graph} (UDG) if there exists a mapping $p : V \to \mathbb{R}^2$ such that, for all distinct vertices $u, v \in V$,
$ uv \in E \text{ if and only if } \| p(u) - p(v) \|_2 \leq 1$. This normalized point-distance definition is equivalent, after scaling,
to representing the graph as the intersection graph of congruent disks. Unit disk graphs are widely used to model wireless and mobile networks.

The Minimum Semitotal Domination problem asks for a semitotal dominating set of minimum cardinality.
The decision version is NP-complete for unit disk graphs \cite{Rou26}. In the graph-based input model,
the graph is supplied by its adjacency structure and is guaranteed to be a UDG; explicit geometric coordinates are not available.

Our contribution is a deterministic $5$-approximation algorithm in the graph-based input model.
The algorithm constructs a maximal independent set by processing breadth-first-search layers in nondecreasing order.
A key feature of this ordering is that every selected vertex has a selected partner at distance two.
With adjacency-list input, the algorithm runs in $O(n + m)$ time.

\section{Related work}

In computational complexity theory, the dominating set problem is a classical NP-complete problem. The domination
problem and its variations have been extensively studied in the literature. For a comprehensive survey on these topics, see the book \cite{Hay20}.

The concept of the semitotal dominating set was introduced by Goddard et al.\cite{God14} in 2014. The authors in \cite{God14}
established that for any connected graph $G$ with $n \geq 4$ vertices, the semitotal domination number
satisfies $\gamma_{t2}(G) \leq \frac{n}{2}$. Additionally, they showed that if $G$ is a graph with $n$ vertices and maximum degree $\Delta$,
then $\gamma_{t2}(G) \geq \frac{2n}{2\Delta+1}$. The Semitotal Domination Decision problem is NP-complete for general graphs \cite{God14}
and remains NP-complete for several restricted graph classes, including planar graphs, split graphs, and chordal bipartite graphs \cite{Hen19}.

On the algorithmic side, minimum semitotal dominating sets can be found in linear time in trees \cite{God14,Hen18} and block graphs \cite{Hen22}.
A polynomial-time algorithm is known for interval graphs \cite{Hen19}, and a linear-time algorithm is available for strongly chordal graphs \cite{Tri23}.

Recently, Rout and Das \cite{Rou26} demonstrated that the problem is also NP-complete for unit disk graphs and proposed
a 6-factor approximation algorithm with $O(n \log n)$
running time under a geometric input model. Henning et al. \cite{Hen26} subsequently obtained a $5.75$-approximation algorithm
with $O(n^3)$ running time in the graph-based input model. The algorithm presented below improves both the approximation guarantee
and the running time in the latter model.

\section{The approximation algorithm}

Our algorithm's approximation guarantee relies on the relationship between the size of an MIS and the minimum cardinality of a dominating set in UDGs.
The following lemma will be used in our performance analysis.

\vspace{4mm}
\noindent\textbf{Lemma 1} \cite{Mar95}. {\em For any UDG $G$, if $I$ is a maximal independent set, then $|I| \leq 5 \cdot \gamma(G)$}.
\vspace{4mm}

A tree is an acyclic connected graph. A {\em Breadth-First-Search} (BFS) tree is a spanning tree constructed by breadth-first search.
It stores the shortest paths from a starting vertex to all other vertices in an unweighted graph,
with vertices organized layer by layer according to their distance from the source.
Every vertex except the root has exactly one parent vertex via which it is first visited.
The collection of all these parent-child edges forms the BFS tree.

Given a graph $G=(V,E)$, select a vertex $r\in V$.
Denote by  $T_{\text{BFS}}$ the BFS tree rooted at $r$ in $G$.
Let $d(r,v)$ denote the distance from vertex $v$ to the root $r$ in $T_{\text{BFS}}$.
After fixing the root $r$, the maximum distance from $r$ to all other vertices
can be easily computed, denoted as $R$.
For each integer $0\leq i\leq R$, define the $i$-th layer of $T_{\text{BFS}}$ as $L_i = \{v\in V \mid d(r,v)=i\}$.
The family $\{L_0,L_1,\dots,L_R\}$ forms a partition of the vertex set $V$, and every vertex in $L_i$ has a parent in $L_{i-1}$,
 $1\leq i\leq R$. Obviously, $L_0 = \{r\}$, $L_1 = N(r)$.

An illustrative example with $T_{\text{BFS}}=3$ is given in Figure 1. The MIS consists of vertex $r$ and five
black vertices; four grey nodes are their parent vertices.

\begin{figure}
		\centering
		\includegraphics[width=0.5\linewidth]{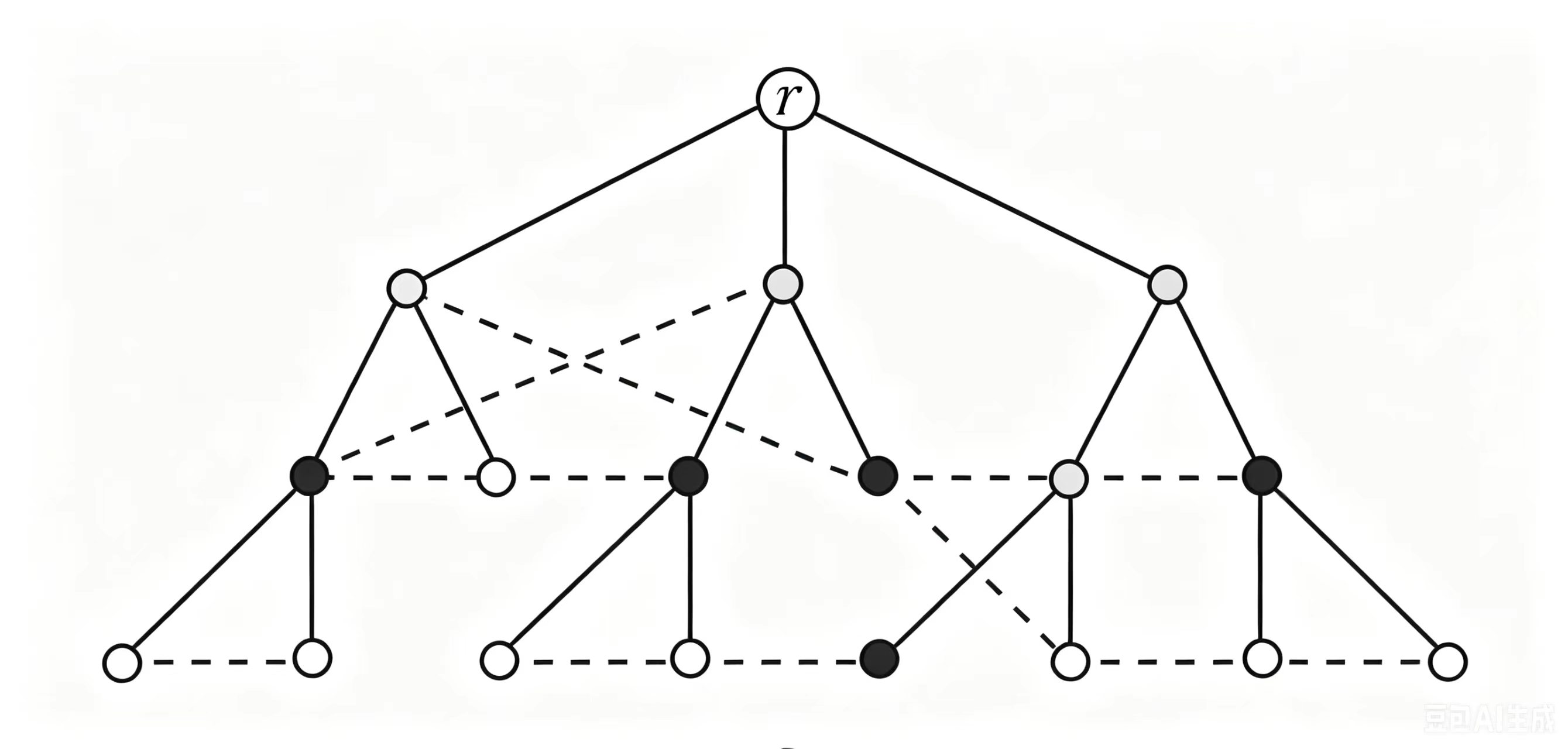}
		\caption{$T_{\text{BFS}}$ consists of solid edges while other edges in $G$ are
dashed}
	\end{figure}

It is well-known that any maximal independent set is a dominating set. In fact, the maximal independent set obtained
via a specific construction method can also be a semitotal dominating set. We now use a greedy algorithm to find a maximal independent set $S$.
The algorithm begins with the root. It then scans the remaining layers in increasing order. Whenever an vertex is encountered,
the vertex is added to the solution and its closed neighborhood is deleted. The only exceptional case is $R=1$, in which the root
together with any one of its neighbors is then optimal.

If $R=1$, choose any vertex $x\in L_1$ and set $S=\{r, x\}$. The set $S$ is a semitotal dominating set.
Every semitotal dominating set has at least two vertices, so $S$ is optimal in this case.
In the following, we assume $R\geq 2$. The algorithm is more formally presented as follows.\\

\vskip2mm

\hrule

\vskip1mm

\noindent{\bf Algorithm 1}  Layer-wise semitotal domination
 \vskip0.4mm\hrule
\vspace{4mm}
\noindent\textbf{Input}: A connected UDG $G$, a root $r$, and its BFS layers $L_1,L_2,\dots,L_R$\\
\noindent\textbf{Output}: Maximal independent set $S$
\begin{enumerate}
\item set $S := \{r\}$; $V' := V \setminus N_G[r]$

\item {\bf for} $i:=2$ {\bf to} $R$

\item \hspace{0.6cm}{\bf while} $L_i \cap V' \neq \emptyset$
\item \hspace{1.2cm} Choose a vertex $v\in L_i\cap V'$

\item \hspace{1.2cm} $S:=S\cup \{v\}$

\item \hspace{1.2cm} $V':=V'\setminus N_G[v]$

\item \hspace{0.6cm} {\bf end while}

\item {\bf end for}

\item {\bf return} $S$

\end{enumerate}
\hrule \vskip 2mm

\vspace{4mm}
\noindent\textbf{Theorem 2} {\em  Let $G = (V,E)$ be a unit disk graph. The set $S$ returned by Algorithm 1
is a semitotal dominating set, and $|S|\leq 5 \cdot\gamma_{t2}(G)$. The algorithm runs in $O(m+n)$ time.}

\noindent\textbf{Proof.} Whenever a vertex is added to $S$, its closed neighborhood is delected. Hence no subsequently selected vertex is adjacent to it,
and $S$ is independent. The root $r$ dominates $L_0\cup L_1$. Every vertex in a later layer is either selected when it is scanned or has already
been deleted by a selected neighbor. Thus $S$ dominates $G$. Therefore $S$ is a maximal independent set.

It remains to verify the semitotal condition. Since $R \ge 2$, the layer $L_2$ is nonempty, at least one vertex $x \in L_2$ is selected.
Consequently, $d(r, x) = 2$, and the root has a partner in $S$.

Now let $u \in S \cap L_i$ for some $i \ge 2$, and let $w \in L_{i-1}$ be the parent of $u$ in the BFS tree. If $i = 2$, then $w \in L_1$
is adjacent to $r$, and hence $d(u, r) = 2$. Suppose that $i \ge 3$. The layer $L_{i-1}$ was completely processed before $u$ was selected.
The parent $w$ cannot belong to $S$, because then selecting $w$ would delete $u$. Hence $w$ must be adjacent to some vertex $v\in S$ in
the same layer or the previous layer, which implies $d(u,v)=2$. Thus every vertex in $S$ has another vertex of $S$ within distance two.
In addition, any maximal independent set is also a dominating set,
so $S$ is a semitotal dominating set. By Lemma 1 and the inequality $\gamma(G)\leq \gamma_{t2}(G)$, we have $|S| \leq 5\cdot\gamma(G)\leq 5 \cdot\gamma_{t2}(G)$.
Thus the approximation factor is at most 5.

Constructing BFS tree and computing a maximal independent set $S$ by the greedy algorithm both take $O(m+n)$ time. Hence, the total time complexity is $O(m+n)$,
which is $O(n^2)$ in the worst case. This completes the proof.
\hfill$\square$

\section*{Acknowledgements}
%The authors would like to thank the anonymous referees for their helpful comments on improving the representation of the paper.

This work is  supported by the National Natural Science Foundation of China under grant number 12571381.

\end{document}